\documentclass[twocolumn,aps,pra,nofootinbib,groupedaddress,amsmath,amssymb,longbibliography]{revtex4-1}
\usepackage{graphicx}
\usepackage{bm}
\usepackage[colorlinks=true,	linkcolor=blue,urlcolor=blue,anchorcolor=blue,citecolor=blue,bookmarksnumbered]{hyperref}
\usepackage{float}

\begin{document}
\title{\textbf{Observation of Blackbody Radiation Enhanced Superradiance in Ultracold Rydberg Gases}}
\author{Liping Hao$^{\star,1,5}$}
\author{Zhengyang Bai$^{\star,2}$}
\email{zhybai@lps.ecnu.edu.cn}
\author{Jingxu Bai$^{1,5}$}
\author{Suying Bai$^{1,5}$}
\author{Yuechun Jiao$^{1,5}$}
\author{Guoxiang Huang$^{2,4,5}$}
\author{Jianming Zhao$^{1,5}$}
\email{zhaojm@sxu.edu.cn}
\author{Weibin Li$^3$}
\email{weibin.li@nottingham.ac.uk}
\author{Suotang Jia$^{1,5}$}
\affiliation{$^{1}$State Key Laboratory of Quantum Optics and Quantum Optics Devices, Institute of Laser Spectroscopy, Shanxi University, Taiyuan 030006, China\\
$^2$State Key Laboratory of Precision Spectroscopy, East China Normal University, Shanghai 200062, China\\
$^3$School of Physics and Astronomy and Centre for the Mathematics and Theoretical Physics of Quantum Non-equilibrium Systems, University of Nottingham, Nottingham, NG7 2RD, UK\\
$^4$NYU-ECNU Joint Institute of Physics, New York University Shanghai, Shanghai 200062, China\\
$^5$Collaborative Innovation Center of Extreme Optics, Shanxi University, Taiyuan, Shanxi 030006, China}
\date{\today}

\def\thefootnote{$\star$}\footnotetext{These two authors contributed equally to this work.}\def\thefootnote{\arabic{footnote}}

\begin{abstract}
An ensemble of excited atoms can synchronize emission of light collectively in a process known as superradiance when its characteristic size is smaller than the wavelength of emitted photons. The underlying superradiance depends strongly on electromagnetic (photon) fields surrounding the atomic ensemble. High mode densities of microwave photons from $300\,$K blackbody radiation (BBR) significantly enhance decay rates of Rydberg states to neighbouring states, enabling superradiance that is not possible with bare vacuum induced spontaneous decay. Here we report observations of the superradiance of ultracold Rydberg atoms embedded in a bath of room-temperature photons. The temporal evolution of the Rydberg $|nD\rangle$ to $|(n+1)P\rangle$ superradiant decay of Cs atoms ($n$ the principal quantum number) is measured directly in free space. Theoretical simulations confirm the BBR enhanced superradiance in large Rydberg ensembles. We demonstrate that the van der Waals interactions between Rydberg atoms change the superradiant dynamics and modify the scaling of the superradiance. In the presence of static electric fields, we find that the superradiance becomes slow, potentially due to many-body interaction induced dephasing. Our study provides insights into many-body dynamics of interacting atoms coupled to thermal BBR, and might open a route to the design of blackbody thermometry at microwave frequencies via collective, dissipative photon-atom interactions.
\end{abstract}

\maketitle

\section{Introduction}
Superradiance describes cooperative radiation of an ensemble of dense excited atoms, in which atomic decay is synchronized collectively by vacuum photon fields. Superradiance leads to faster and stronger light emission than independent radiations. Since predicted by Dicke in 1954~\cite{dicke_coherence_1954}, superradiance has been observed in a variety of systems~\cite{Gross_Maser_1979,Goy_Rydberg_1983,devoe_observation_1996,eschner_light_2001,wang_superradiance_2007,scheibner_superradiance_2007,Rohlsberger_Lamb_10,keaveney_cooperative_2012,Meir_Lamb_ion_2014,mlynek_observation_2014,goban_superradiance_2015,araujo_superradiance_2016,roof_observation_2016,chen_experimental_2018,Wen_Lamb_Superconducting_2019,haber_spectral_2019}. Superradiance plays important roles in understanding fundamentally important light-matter interactions and phase transitions~\cite{Baumann_Dicke_2010,Zhu_Squeezed_2020}. Recently, it has been shown that superradiance finds applications in realizing quantum metrology~\cite{Wang_Heisenberg_2014,Liao_Gravitational_2015}, laser~\cite{Haake_Superradiant_1993,bohnet_steady-state_2012,Svidzinsky_Quantum_2013,zhu_synchronization_2015-1,Norcia_Cold_2016}, and atomic clocks~\cite{Norcia_Superradiance_2016}, etc..
\begin{figure}
	\centering
	\includegraphics[width=1\linewidth]{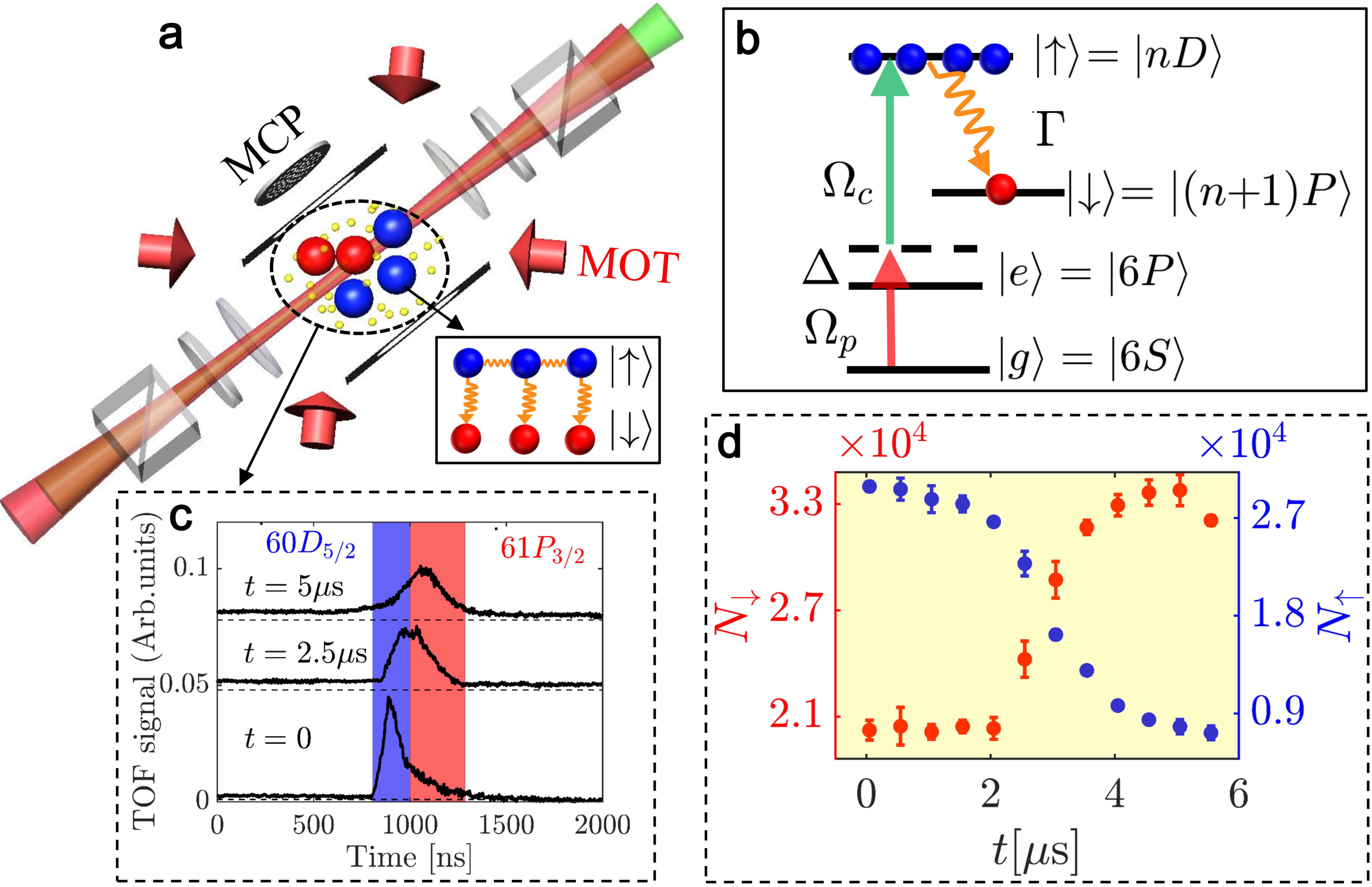}
	\caption{\footnotesize(Colour online)  \textbf{Rydberg superradiance of caesium atoms.} (a) Experimental setup. The coupling laser $\lambda_{c}$ = 510~nm and probe laser $\lambda_{p}$ = 852~nm counter-propagate through the trap center. At time $t$,  Rydberg atoms are field ionized and detected at the MCP. (b) Two-photon Rydberg excitation. The probe light (Rabi frequency $\Omega_{p}$) drives the lower transition $|6 S_{1/2}, F=4\rangle $ $\to$ $|6P_{3/2}, F'=5\rangle $, and is blue detuned 360 MHz from $|6P_{3/2}, F'=5\rangle$ using a double-pass acousto-optic modulator (AOM). The control light (Rabi frequency $\Omega_c$) couples the transition $|6P_{3/2}, F'=5 \rangle $ $\to$ $|\uparrow\rangle=|n D_{5/2}\rangle$. The $|\uparrow\rangle$ state decays to a neighbouring Rydberg state $|\downarrow\rangle=|(n+1)P_{3/2}\rangle$ at decay rate $\Gamma\sim\,$kHz. (c) Snapshots of ion signals. The first gate (blue) measures populations in state $|\uparrow\rangle=|60D_{5/2}\rangle$. The second gate (red) gives populations in state $|\downarrow\rangle=|61P_{3/2}\rangle$. (d) Evolution of Rydberg atom number $N_{\downarrow}$ ($N_{\uparrow}$). The dynamics is slow when $t<2\,\mu$s and accelerated rapidly when $t>2\,\mu$s. $N_{\downarrow}$ ($N_{\uparrow}$) reaches the maximal (minimal) value at around $t=5\,\mu$s. This time scale is much shorter than the lifetime of Rydberg atoms $\sim$ms.
MOT: magneto-optical trap. MCP: microchannel plate.}\label{model}
\end{figure}

When the surrounding photon field is modified locally by, e.g. cavities, characters of light-matter interactions change drastically, leading to unconventional phenomena such as the paradigmatic Casimir ~\cite{PhysRevLett.78.5} and Purcell effects~\cite{Purcell,Purcell_exp}. A thermal bath of blackbody photons can modify the interaction, too. This causes tiny energy shifts to groundstate atoms, and can  be detected by accurate optical clocks~\cite{PhysRevLett.91.173005,bloom_optical_2014}. In electronically high-lying Rydberg states, atoms can strongly interact with blackbody radiation (BBR)~\cite{Archimi_Measurements_2019}. At room temperature $T$, BBR photons of low-frequency microwave (MW) fields can provide successive energies to couple different Rydberg states, i. e. $kT>\hbar \omega$  ($k$, $\hbar$, and $\omega$ to be the Boltzmann constant, Planck constant and transition frequency). Due to high numbers of MW photons per mode in the BBR field~\cite{cantatmoltrecht_2020_longlived}, the decay of single Rydberg atoms is orders of magnitude faster than in vacuum. The increase of decay rates~\cite{Gallagher_Interactions_1979,Archimi_Measurements_2019} and energy shifts~\cite{Hollberg_shift} of Rydberg atoms have been measured. The large wavelength ($\sim$mm) of MW photons moreover permit superradiance of Rydberg atoms. Superradiance of Rydberg atoms driven by vacuum fields has been reported \cite{wang_superradiance_2007}. However, Rydberg superradiance induced by thermal BBR has only been observed in the presence of cavities~\cite{Raimond_collective_1982}.

In this work, we report the observation of the superradiance of high-lying Rydberg $|nD_{5/2}\rangle$ states of caesium atoms in a magneto-optical trap (MOT), triggered by room-temperature BBR. Superradiance of the Rydberg atom ensembles is induced by thermal MW photons of wavelength $\sim\,$mm, which is much larger than spatial extensions $\sim\,\mu$m of the atomic gases~\cite{dicke_coherence_1954}.  We measure the superradiant decay in selective $|nD_{5/2}\rangle \to |(n+1)P_{3/2}\rangle$ transition, and scaling with respect to atom numbers and Rydberg states. We identify that the superradiant decay is strongly influenced by van der Waals (vdW) interactions of Rydberg atoms, confirmed by careful theoretical analysis and large-scale numerical simulations. Our study opens a window to experimentally explore the superradiant dynamics of interacting many-body systems coupled to thermal BBR, and enable to develop blackbody thermometry at microwave frequencies through collective photon-Rydberg atom interactions.

The remainder of the article is arranged as follows. In Sec. II, the experiment setup is described and the fast Rydberg decay between $|nD\rangle$ and $|(n+1)P\rangle$ transitions observed in the experiment is presented. In Sec. III, the lifetime of the Rydberg states is estimated and the superradiance dependence on the MW wavelength, atomic number, and BBR temperature is discussed theoretically. In Sec. IV, a master equation model
including van der Waals interactions between Rydberg atoms is
introduced and the mean field simulation on the master equation is carried out,
with the theoretical result compared with the experimental one. In Sec. V, the scaling of the Rydberg superradiance is calculated, and the dependence of the superradiance on the particle number, Rydberg states, and BBR temperature are provided both experimentally and theoretically.
In Sec. VI, a preliminary experimental result on the superradiance dynamics  in the presence of MW and static electric fields is given, which manifests the signature of dipole-dipole interactions between atoms. Lastly, Sec. VII contains a summary of the research results obtained in this work.

\section{Experiment}
In our experiment, up to $10^7$ caesium atoms are laser cooled to $100\,{\rm \mu K}$ and trapped in a spherical (diameter $\approx 550\,\mu $m) MOT~[see Fig.~\ref{model}(a)]. Starting from the groundstate $|6 S_{1/2}, F=4\rangle$, Rydberg $|\uparrow\rangle=|nD_{5/2}\rangle$ state is excited through intermediate state $|6P_{3/2},F'=5\rangle$. The level scheme is depicted in Fig.~\ref{model}(b). Both the probe and control lasers are linearly polarized and counter propagating through the MOT center (with corresponding waists $80\,\mu$m and $40\,\mu$m) forming a cylindrical excitation region.

In each experiment cycle, atoms are excited to the Rydberg state ($n\geq 60$) in $6\,\mu$s (after turning off the trap laser). Due to blockade by the vdW interaction (blockade radius $R_b$), number $N_e$ of Rydberg atoms is varied between $10^3$ to $10^4$ by changing the laser power. After switching off the excitation laser, Rydberg atoms are allowed to evolve for a duration $t$, and then ionized by a state-dependent electric field. The ions are detected by a microchannel plate (MCP) detector with efficiency about $10\%$. The detail of our experiment can be found in Appendix~\ref{exp}.

\begin{figure}
	\centering
	\includegraphics[width=1\linewidth]{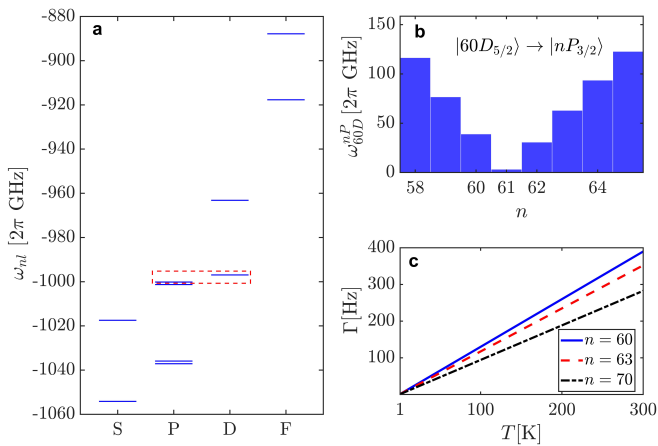}
	\caption{\footnotesize(Colour online) \textbf{Properties of caesium atoms}. (a) Energy levels $\omega_{nl}$ of caesium atoms. The principal quantum numbers are $n=60, \text{and} \,61$. State $|60D_{5/2}\rangle$ and $|61P_{3/2}\rangle$ are energetically close.  (b)  The $|60D\rangle \to |nP\rangle$ transition frequency $\omega_{60D}^{nP}$, the lowest transition frequency $\omega_{60D}^{61P}$ $\simeq2\pi\times3.2\,$GHz.
		(c) The decay rate from $|nD\rangle$ state to $|(n+1)P_{3/2}\rangle$ for different temperature $T$ (with $n=60, 63$ and $70$). At room temperature, the decay rate is significantly increased.}\label{MW_Coupling}
\end{figure}

The state-selective ionization and detection method shows that Rydberg state $|nD_{5/2}\rangle$ decays immediately to the energetically closest $|(n+1)P_{3/2}\rangle$ state [see example of atomic levels in Fig.~\ref{MW_Coupling}(a)]. In Fig.~\ref{model}(c), snapshots of ion signals are shown for state $|60D_{5/2}\rangle$. Increasing time $t$, the population transfers to $|61P_{3/2}\rangle$ state rapidly where the peak is drifted towards a later time. The population of $|61P_{3/2}\rangle$ state is obtained in the red gate region. The tail in this gate when $t=0$ indicates that the decay occurs slightly also during the Rydberg excitation.

The population dynamics of the system displays qualitatively different behaviors at later time. As shown in Fig.~\ref{model}(d),  populations change slowly when $t<2\,\mu$s. During $2\,\mu \text{s} < t < 4\,\mu$s,  a large portion of the population is transferred to state $|61P_{3/2}\rangle$ rapidly. The population in state $|61P_{3/2}\rangle$ reaches maximal at $t\approx 5\,\mu$s, and then decays to other states when $t>5\,\mu$s. Such fast decay, much shorter than the lifetime $2.6\,$ms  in the underlying transition at room temperature, is rooted from superradiance of the Rydberg ensemble interacting with BBR photons. For convenience, quantum number $J=5/2$ and $J'=3/2$ will be omitted in the notation from now on.

\section{BBR enhanced Rydberg superradiance}
Decay of Rydberg atoms is affected by BBR and such effect has been experimentally observed. For example, the recent experiment has found that lifetimes in Rydberg $nS$ state are determined by 300K BBR~\cite{PhysRevLett.124.023201}.  To identify the lifetime between Rydberg levels,
the decay rate of spontaneous transition between $nJ$ and $n^\prime J^\prime$ states can be calculated by~\cite{Beterov_Quasiclassical_2009, Gallagher_Interactions_1979,Theodosiou_Na_sequence_1998},
\begin{eqnarray} \label{EinsteinJ}
\Gamma_{nJ}^{n^\prime J^\prime}=\frac{{\omega_{nJ}^{n'J'}}^3}{2\pi\epsilon_0\hbar m_ec^3}\frac{2J+1}{2J^\prime+1}|\langle nJ|e\mathbf{r}|n^\prime J^\prime \rangle|^2,
\end{eqnarray}
where $m_e$ is the electron mass, $\omega_{nJ}^{n'J'}=|E_{nJ}-E_{n^\prime J^\prime}|$ is the transition frequency, with $E_{nJ}$ and $E_{n^\prime J^\prime}$ being energies of $nJ$ and $n^\prime J^\prime$ states, respectively. Energies $E_{nL}=-1/(2n_{\rm eff}^2)$ [in atomic units] of the Rydberg states are expressed through the effective quantum number $n_{\rm eff}=n-\mu_J$, where $\mu_J$ is a quantum defect of Rydberg $nJ$-state, which can be found in Ref.~\cite{Deig_2016}.

The lifetime of a Rydberg state depends on background BBR temperature. Taking into account of photon number per mode at temperature $T$, the decay rate becomes,
$\Gamma_{nJ}^{n^\prime J^\prime}(T) = \Gamma_{nJ}^{n^\prime J^\prime}\bar{n}_\omega(T),$
where the thermal factor $\bar{n}_\omega(T)=1/[{\rm exp}(\hbar\omega_{nJ}^{n'J'}/k_BT)-1]$  gives Bose-Einstein statistics of photon numbers at temperature $T$. The total decay rate is $\Gamma_{nJ}(T) = \sum_{n'J'} \Gamma_{nJ}^{n'J'}(T)$~\cite{Beterov_Quasiclassical_2009,ovsiannikov_rates_2011}.
For MW transitions, the photon energy is far smaller than the thermal energy, i.e. $\hbar \omega_{nn'}\ll k_BT$, such that the thermal factor is far larger than 1.

As an example, we show that the transition frequency between $|60D\rangle$ and $|61P\rangle$ state in  Fig.~\ref{MW_Coupling}(a). The transition frequency $\omega_{60D}^{61P}\approx2\pi\times3.2$\,GHz is far smaller than other transition energy [highlighted with a box]. Though MW photons of various frequencies can be emitted with higher rates, superradiance on the other hand enhances emission rates of selected transitions, depending on MW wavelengths $\lambda_{nl}^{n'l'}$.  Wavelengths of the MW photon corresponding to $|60D\rangle\to |61P\rangle$ is about $92.93$ mm $\gg d=550 \,\mu$m (the size of the atomic sample)  or wavelengths of other MW photons. We also plot the decay rate $\Gamma$ from $|nD\rangle$ state to $|(n+1)P_{3/2}\rangle$ as functions of temperature $T$ in Fig.~\ref{MW_Coupling}(c). Here $\Gamma$ almost linearly increase with $T$. Clearly at room temperature, the decay rate is greatly enhanced by $\bar{n}_\omega(T)\gg 1$. Superradiance in the $|nD\rangle\to|(n+1)P\rangle$ transition is much stronger than one of the $|nS\rangle \to |nP\rangle$ transition~\cite{wang_superradiance_2007}, due to the low frequency MW transition.
\begin{figure}
	\centering
	\includegraphics[width=1\linewidth]{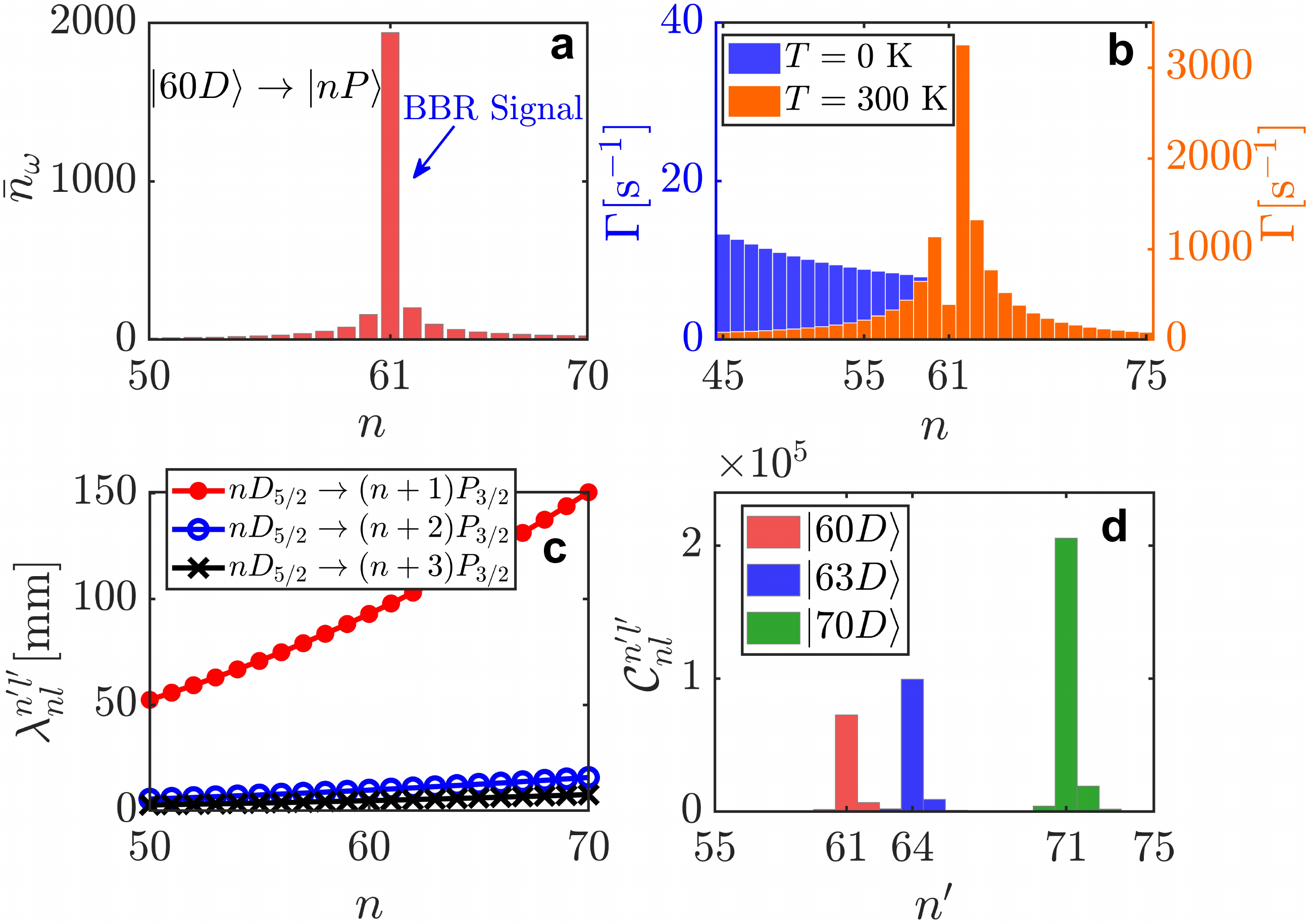}
	\caption{\footnotesize(Colour online) \textbf{Calculating results of the superradiance parameters.}
		(a) BBR photon number $\bar{n}_\omega(T)$ in  $|60D\rangle\to|nP\rangle$ transition at $T=300$K. The maximal value of $\bar{n}_\omega(T)$ is found in the transition $|60D\rangle\to|61P\rangle$. (b) Decay rate of $|60D\rangle \to |nP\rangle$ transition at $T=0$ (blue) and  $T=300$K (orange). As  $\Gamma_{nl}^{n'l'}(0)\propto (\omega_{nl}^{n'l'})^3$, the rate becomes larger when decaying to lower states at $T=0$. At $T=300$K, BBR enhances decay rates corresponding to MW transitions. (c) Transition wavelength for different $|n D_{5/2}\rangle\to |(n+a)P_{3/2}\rangle$ transition $(a=1, 2, 3)$. The $|nD_{5/2}\rangle\to |(n+1)P_{3/2}\rangle$ transition gives the largest wavelength (tens of mm), far larger than the spatial dimension of the trap, enabling superradiant decay. (d) Superradiance threshold parameter at $T=300$K.  Parameter $\mathcal{C}_{nl}^{n'l'}$ in the $|n D\rangle \to |(n+1)P\rangle$ transitions is orders of magnitude larger than other transitions, due to large wavelengths and high mode densities of the MW photons.}\label{thermal_effect}
\end{figure}

\begin{figure*}
	\centering
	\includegraphics[width=0.9\linewidth]{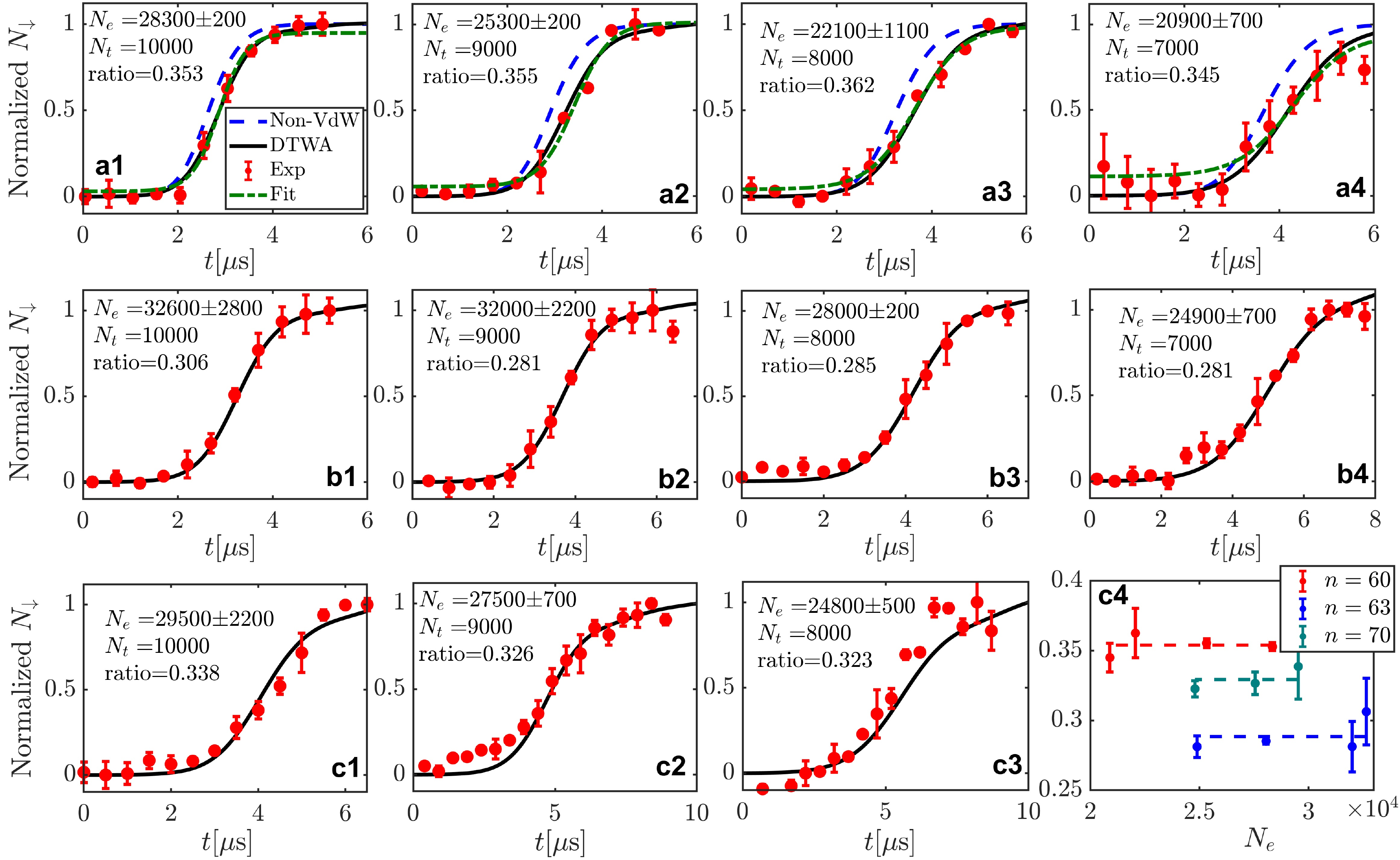}
	\caption{\footnotesize(Colour online) \textbf{Population evolution.} (a1)-(a4) Net changes of the $|61P\rangle$ population.  Experimental data (dot) and master equation simulation (solid) agree well. The green dash-dotted curve corresponds to fitting the experimental data. The blue dashed curve is the analytical Eq.~(\ref{eq:freen}). The ratio $N_t/N_e\approx 0.35$ is largely independent of data sets. (b1)-(b4) net changes of the $|64P\rangle$ population.  The ratio $N_t/N_e\approx 0.28$ for different data sets. (c1)-(c3) net changes of the $|71P\rangle$ population, and  the ratio $N_t/N_e\approx 0.32$ for different data sets. (c4) The ratio $N_t/N_e$ for different principal quantum number $n$. The ratio stays a constant when varying $N_e$ at a given $n$, showing consistency of the experiment and simulation.}\label{Superradiance}
\end{figure*}
In the experiment,  frequency  $\omega_{60D}^{61P}\approx 2\pi\times3.2\,$GHz in $|60D\rangle \to |61P\rangle$ transition, far smaller than $k_B T/\hbar=2\pi\times 6248\,$GHz when $T=300\,$K. The corresponding photon number approaches 2000 [Fig.~\ref{thermal_effect}(a)], which amplifies the underlying decay rate  by three orders of magnitude [Fig.~\ref{thermal_effect}(b)].
Very importantly, superradiance is triggered in this transition, as the wavelength of the MW photons is 92.93 millimeters [Fig.~\ref{thermal_effect}(c)], much larger than the dimension of the gas. The strength of superradiance is characterized by a threshold parameter $\mathcal{C}_{nl}^{n'l'}=\Gamma_{nl}^{n'l'}(T)\mathcal{G}_{nl}^{n'l'}/\Gamma_{nl}(T)$, where $\mathcal{G}_{nl}^{n'l'}=3(\lambda_{nl}^{n'l'})^2/8\pi d^2$ is a form factor and $\lambda_{nl}^{n'l'}$ is the wavelength of underlying transition~\cite{Gross_Maser_1979}.  The larger the parameter $\mathcal{C}_{nl}^{n'l'}$ is, the stronger superradiance takes place. As shown in Fig.~\ref{thermal_effect}(d), the threshold parameter $(\sim 10^5)$ corresponding to the $|60D\rangle\to |61P\rangle$ transition is several orders of magnitude larger than that of other transitions~\cite{wang_superradiance_2007}.

The resulting strong superradiance exhibits sensitive dependence on numbers of the Rydberg atoms. In Fig.~\ref{Superradiance}(a)-(c), net changes of the $|61P\rangle$ population, i.e. growth of the Rydberg atom number $N_e$ when  $t\ge0$, are shown. A generic feature is that populations increase rapidly and arrive at maximal values, after a slow varying stage. Increasing $N_e$, the population dynamics become faster such that it takes less time to reach the maxima.


\section{Master equation simulation}
Dipole-dipole interactions between Rydberg atoms can slow down or even destroy superradiance~\cite{Grimes_Direct_2017}. In our experiment, the dipole-dipole interaction plays a negligible role in the decay, where the angular average of the dipolar interactions  vanishes in such large ensemble.

For vanishing dipole-dipole interaction and large single-photon detuning, the dynamics of the superradiant decay can be modeled by the quantum master equation for the many-atom density operator $\rho$:
\begin{eqnarray} \label{master}
\frac{\partial\rho}{\partial t}=-i[H,\rho]+D(\rho),
\end{eqnarray}
under a two-level approximation. The Hamiltonian in the equation is given by
\begin{eqnarray} \label{Hami}
	H&=&
	\sum_{k\neq j}^{N_t} \hbar \left[\frac{1}{2}V_{D}(\mathbf{r}_{jk})\hat{n}^j_{\uparrow\uparrow}
\hat{n}^k_{\uparrow\uparrow} + \frac{1}{2} V_{P}(\mathbf{r}_{jk})\hat{n}^j_{\downarrow\downarrow}
\hat{n}^k_{\downarrow\downarrow}\right].
\end{eqnarray}
where $\hat{n}^j_{\uparrow\uparrow}=\hat{\mathbb{I}}/2-\hat{S}_z^j$ and $\hat{n}^j_{\downarrow\downarrow}=\hat{\mathbb{I}}/2+\hat{S}_z^j$  ($\hat{\mathbb{I}}$ is identity operator) are atomic number operators respectively at the upper state  $|\uparrow\rangle$ and the lower state $|\downarrow\rangle$, $V_{D(P)}=C_6^{D(P)}/|{\bf r}_j-{\bf r}_k|^6$ is the vdW potential with the dispersive coefficient $C_6^{D(P)}\propto n^{11}$. Note that influences of the vdW interaction on superradiance have not been explored so far.

In Eq.~(\ref{master}) the radiative decay from the upper state  $|\uparrow\rangle$ to the lower state $|\downarrow\rangle$ is described by collective dissipation of Lindblad form~\cite{FICEK_Entangled_2002}
\begin{eqnarray} \label{Lindblad1}
D(\rho)=&&\sum_{j,k=1}^{N_t}\hbar \Gamma\left(\hat{S}_{-}^j\rho\hat{S}_{+}^k-\frac{1}{2}\{\hat{S}_{+}^k\hat{S}_{-}^j, \rho\}\right).
\end{eqnarray}
where $\Gamma=\Gamma_{nD}^{(n+1)P}(T)$ is the decay rate, whose spatial dependence can be neglected as averaging spacing between Rydberg atoms is much smaller than $\lambda_{nD}^{(n+1)P}$. In both (\ref{Hami}) and (\ref{Lindblad1}), ${\mathbf S}_j = (\hat{S}_x^j, \hat{S}_y^j, \hat{S}_z^j)$ are the Pauli matrix of the $j$th atom, with raising and lowering operator $\hat{S}_\pm^k=\hat{S}_x^k\pm i\hat{S}_y^k$.

For small systems (i.e., a few tens of atoms), the quantum master equation can be solved  by direct diagonalization.
However, the number of Rydberg excitation is large (i.e., $10^3\sim10^4$) in experiment.  To efficiently simulate a large system (total particle number $N_t\gg 1$), one can apply the method of the discrete truncated Wigner approximation (DTWA), which is a phase space method by which the density-operator equation can be replaced by its mean-value equation with the quantum fluctuations of the system involved in random initial states~\cite{Schachenmayer_DTWA_2015, Schachenmayer_2DDTWA_2015}.

Based on the idea of the DTWA, we define mean values ${\mathbf s}_k=\langle{\mathbf S}_k\rangle$ for our system. Then we obtain the equations of motion of ${\mathbf s}_k$  associated with the master equation (\ref{master}), with the form
\begin{subequations}\label{TWOLevel}
\begin{eqnarray}
\frac{\partial{s}_{x}^k}{\partial t}=&&-{s}_{z}^k\sum_{j=1}^N\Gamma_{jk}{s}_{x}^j-{s}_{y}^k\sum_{j=1,j\neq k}^N\frac{V_{11}}{2}(0.5+{s}_{z}^j)\nonumber\\
&&+{s}_{y}^k\sum_{j=1,j\neq k}^N\frac{V_{22}}{2}(0.5-{s}_{z}^j),\\
\frac{\partial{s}_{y}^k}{\partial t}=&&-{s}_{z}^k\sum_{j=1}^N\Gamma_{jk}{s}_{y}^j+{s}_{x}^k\sum_{j=1,j\neq k}^N\frac{V_{11}}{2}(0.5+{s}_{z}^j)\nonumber\\
&&-{s}_{x}^k\sum_{j=1,j\neq k}^N\frac{V_{22}}{2}(0.5-{s}_{z}^j),\\
\frac{\partial{s}_{z}^k}{\partial t}=&&\sum_{j=1}^N\Gamma_{jk}({s}_{x}^j{s}_{x}^k+{s}_{y}^j{s}_{y}^k),
\end{eqnarray}
\end{subequations}
In the DTWA method, we describe the initial state by a Wigner probability distribution, $p_{\mu, a_{\mu}}^{k} (\mu=x, y, z;$~the subscript $a_{\mu}$ denotes the index of each trajectory, $k$ denotes the position of Rydberg atom) for certain discrete configurations of Bloch vector elements, $s_{\mu}^k=\langle\hat{S}_{\mu}^k\rangle$. Consider the eigen-expansion of the spin operators,  $\hat{S}^{\mu}_k=\sum_{a_{\mu}}\eta_{\mu,a_{\mu}}^{k}|\eta_{\mu,a_{\mu}}^{k}\rangle\langle\eta_{\mu,a_{\mu}}^{k}|$, where $\eta_{\mu,a_\mu}^{k}$ and $|\eta_{\mu,a_\mu}^{k}\rangle$ denote the eigenvalues and eigen-vectors, respectively. Then, we select the ``a-{\rm th}'' eigenvalue, $\lambda_{\mu}^{k}(t=0)=\eta_{\mu,a_\mu}^{k}/2$, with probability $p_{\mu,a_\mu}^{k}={\rm Tr}[\hat{\rho}_0^{k}|\eta_{\mu,a_\mu}^{k}\rangle\langle\eta_{\mu,a_\mu}^{k}|]$. Specifically, all the atoms initially populate in the upper state $|\uparrow\rangle$, with initial density matrix $\hat{\rho}^{k}_0=|\uparrow\rangle\langle\uparrow|$, which leads to fixed classical spin component along $z$, $\sigma_{z}^k=-1/2$, and fluctuating spin components in the orthogonal directions $\sigma_{x(y)}^k\in\{-1/2, 1/2\}$, each with $50\%$ probability. Mean values of observable (i.e., the Rydberg population) are calculated by averaging over many trajectories. In the simulation, we consider an ensemble of Rydberg atoms separated by the blockade radius $R_b$ and with an Gaussian distribution in space. Typically we run $\geq10^4$ trajectories to obtain mean values through the ensemble average, which guarantee the convergence of DTWA results. A generalized truncated Wigner approximation (GDTWA) method is give in the  Appendix \ref{4LS} for spin-3/2 atoms when simulating dynamics involving all four levels~\cite{lepoutre_out--equilibrium_2019}.


\section{Scaling of Rydberg superradiance}
Without vdW interactions, the master equation can be solved analytically, yielding the solution to $N_{\downarrow}$~\cite{gross_superradiance:_1982},
\begin{equation}
	\label{eq:freen}
	N_{\downarrow}=\frac{N_t}{2}+\frac{N_t}{2}{\rm tanh}\left[\frac{\Gamma(N_t+1)}{2}(t-t_d)\right],
\end{equation}
where  $t_d={\rm In}(N_t)/[\Gamma(N_t+1)]$ is the delay time, and the collective decay rate $\propto N_t\Gamma$. The analytical solution $N_{\downarrow}$ predicts faster superradiant transition [dashed curves in Fig.~\ref{Superradiance}(a1)-(a4)].

By taking into account of the vdW interaction, our numerical simulation agrees nicely with the experimental data [Fig.~\ref{Superradiance}(a1)-(a4)]. The slower superradiance can be understood that the vdW interaction mixes superradiant and other states. As the latter typically decay slower, such dephasing therefore increases the superradiant decay time. The number $N_t$ of Rydberg atoms used in the simulation is about $35\%$ of the experimental value $N_e$. This difference could attribute to the fact that only some of the Rydberg atoms in the trap are in superradiant states, as we observe  atoms remain in the initial state even when $t>5\,\mu$s [Fig.~\ref{model}(d)].

We have also plotted dynamical evolution of $N_{\downarrow}$  for state $|63P\rangle$ and $|71P\rangle$ in the Fig.~\ref{Superradiance}(b1)-(b4) and ~\ref{Superradiance}(c1)-(c3). DTWA simulations capture our experimental data very well. In Fig.~\ref{Superradiance}(c4), one sees that the number $N_t$ of Rydberg atoms used in the simulation is about $25\sim 35\%$ of the experimental value $N_e$ for all Rydberg states ($n=60$, $n=63$, and $n=70$). The ratio fluctuates around a constant when increasing $N_e$ for a given Rydberg state, which indicates that the experiment and corresponding simulation are consistent.
The finite detection efficiency of the MCP might affect values of $N_e$, and hence the ratio $N_t/N_e$. The time scale, however, is not affected by the detection efficiency, as the ion signal is linearly proportional to $N_e$.

\begin{figure}
	\centering
	\includegraphics[width=0.45\textwidth]{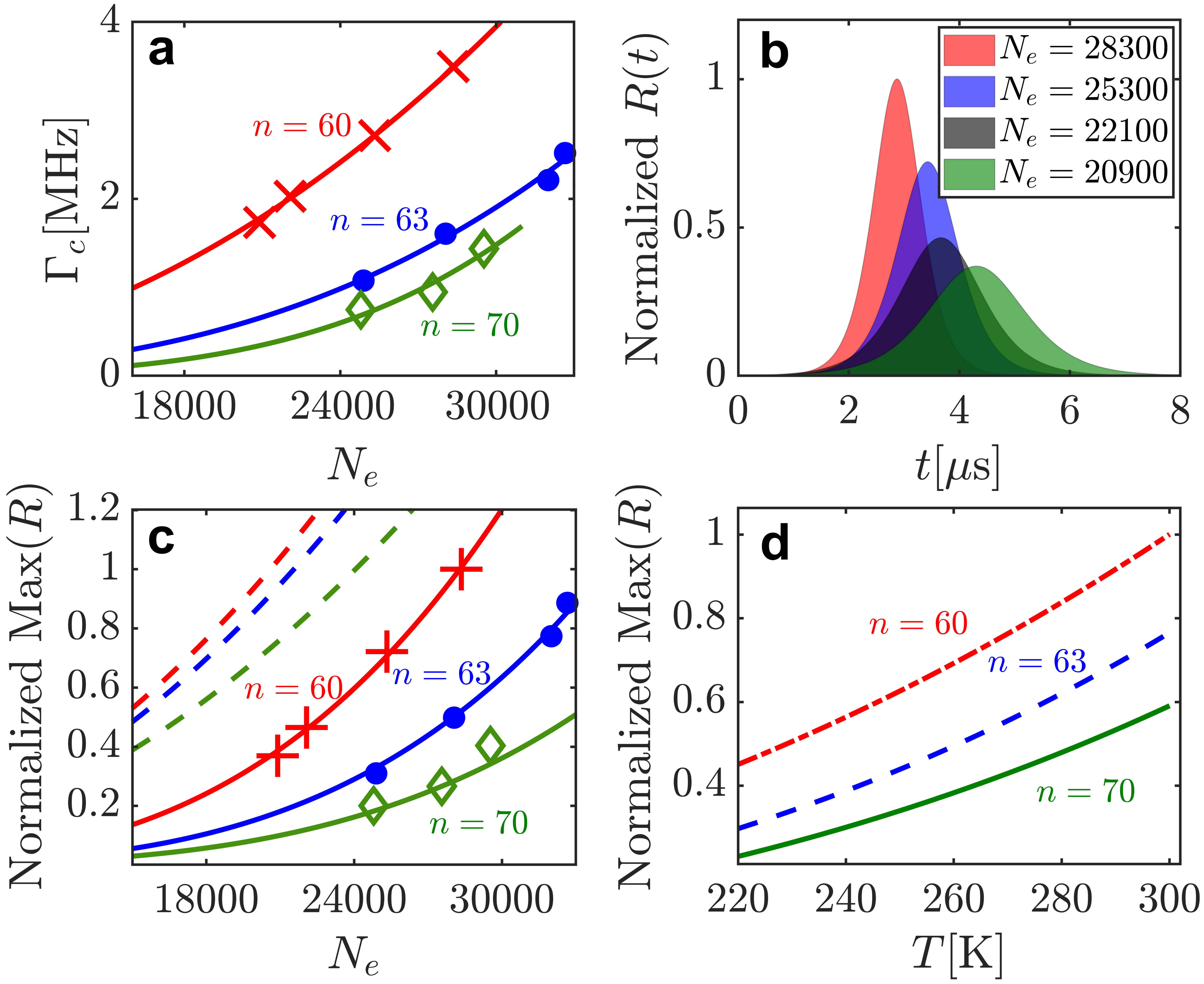}
	\caption{\footnotesize(Colour online) \textbf{Scaling of the Rydberg superradiance.} (a)  Collective decay $\Gamma_c$ obtained from fitting the experiment (labels) and numerical simulations (solid). $\Gamma_c$ is proportional to $N_e^\alpha$. Here $\alpha$ are $2.24$ (red), $2.96$ (blue), and $4.07$ (green) for $n=60$, $63$, and $70$, respectively.  In interaction-free case, $\alpha=1$. (b) Normalized photon emission rate $R(t)$ for different $N_e$. The peak height becomes lower and delay time increases for smaller $N_e$. (c) Maximal rate $R_{\text{max}}$. We find $R_{\text{max}} \propto N_e^\beta$, where $\beta$ are $3.14$ (red), $3.56$ (blue), and $3.62$ (green) for $n=60$, $63$, and $70$, respectively. Dashed lines represent the interaction-free case $\beta=2$ for $n=60$ (red), $63$ (blue), and $70$ (green). (d) Maximal rate $R_{\text{max}}$ as a function of BBR temperature $T$. Here $R_{\text{max}}\propto T^{\xi}$, where $\xi= 2.57, 3.01, 3.12$ for $n=60$, $63$, and $70$. In the simulation, particle number is $N_t=10000$.
	}\label{scaling}
\end{figure}

Drastically, the vdW interaction alters scaling of superradiance with respect to $N_e$ and principal quantum numbers. First, the collective decay rate $\Gamma_c$ changes nonlinearly with $N_e$, which is confirmed by the numerical simulation, shown in Fig.~\ref{scaling}(a). In contrast to the interaction-free case [see Eq.~(\ref{eq:freen})], the rate $\Gamma_c \propto N_e^\alpha$, where $\alpha$ increases from $2.24$ ($n=60$), to $2.96$ ($n=63$) and $4.07$ ($n=70$). Due to stronger vdW interactions ($\propto n^{11}$) and smaller decay rate ($\propto n^{-3}$), the collective rate decreases in higher-lying states.

Next, we study the emission rate of MW photons, given by $r(t)= \dot{N}_{\downarrow}$. Without vdW interactions, the emission rate can be derived from Eq.~(\ref{eq:freen}),
\begin{eqnarray}\label{emission}
r(t)=\frac{\Gamma N_t^2}{4}{\rm sech}^2\left[\frac{\Gamma(N_t+1)}{2}(t-t_d)]\right],
\end{eqnarray}
which has the maximal emission rate $r_{m}=\Gamma N_t^2/4$ at $t=t_d$, i.e. proportional to $N_t$ quadratically.

The emission rate is obtained by fitting the experimental data [green dash-dotted curves in Fig.~\ref{Superradiance}(a1)-(a4)]. In Fig.~\ref{scaling}(b) normalized rate $R(t)=4r(t)/\Gamma N_r^2$ is shown, where $N_r$ is the largest $N_e$ among experiments for a Rydberg state $|nD\rangle$. For example, $N_r=28300$ for state $|60D\rangle$ [see Fig.~\ref{Superradiance}(a) and~\ref{scaling}(a)]. Profiles of
 $R(t)$ exhibit a single peak whose location varies with $N_e$. The maximal value $R_{\text{max}}$ is 1 when $N=N_r$ and $t=t_d$, and smaller than 1 when $N_e<N_r$ [Fig.~\ref{scaling}(b)].

In Fig.~\ref{scaling}(c), $R_{\text{max}}$ as a function of $N_e$ is shown. Both the experimental data and simulation show $R_{\rm{max}}\propto N_e^{\beta}$. Due to strong vdW interactions, the power $\beta$ increases from $3.14$ ($n=60$), to $3.56$ ($n=63$) and $3.62$ ($n=70$). Moreover, the peak rate $R_{\text{max}}$ depends also on the BBR temperature. Our numerical simulations show $R_{\text{max}}\propto T^{\xi}$, where $\xi = 2.57,\, 3.01$, $3.12$ for $n=60,\, 63$ and $70$, respectively [Fig.~\ref{scaling}(d)]. Such dependence might enable a way to measure BBR temperatures.
\begin{figure}
	\centering
	\includegraphics[width=0.80\linewidth]{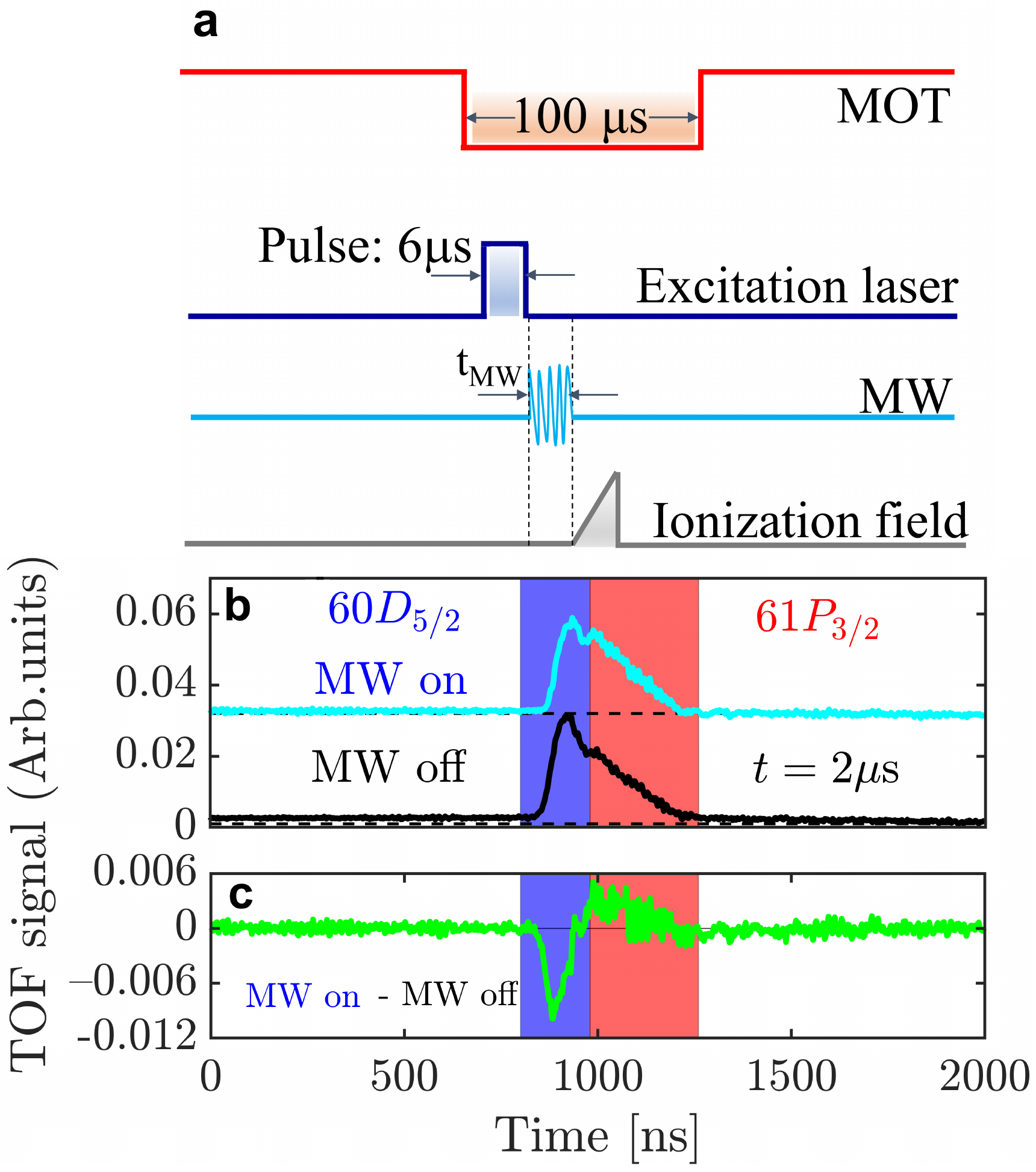}\\
	\caption{\footnotesize \textbf{Superradiance with and without MW fields}. (a) The MW field is applied during the superradiant decay. (b) Population of state $|60D_{5/2}\rangle$ and $|61P_{3/2}\rangle$ at $t=2\mu$s [corresponding to Fig.1(c)]. (c) The population difference with and without MW fields. One sees that the population in state $|60D_{5/2}\rangle$ decreases, and population in state $|61P_{3/2}\rangle$ increases due to the MW field coupling. The MW field modulate the population dynamics dramatically (b). The number of Rydberg atoms is about 26300.}\label{MW}
\end{figure}

\section{Superradiant dynamics with MW and static electric fields}
\begin{figure*}
	\centering
	\includegraphics[width=0.9\linewidth]{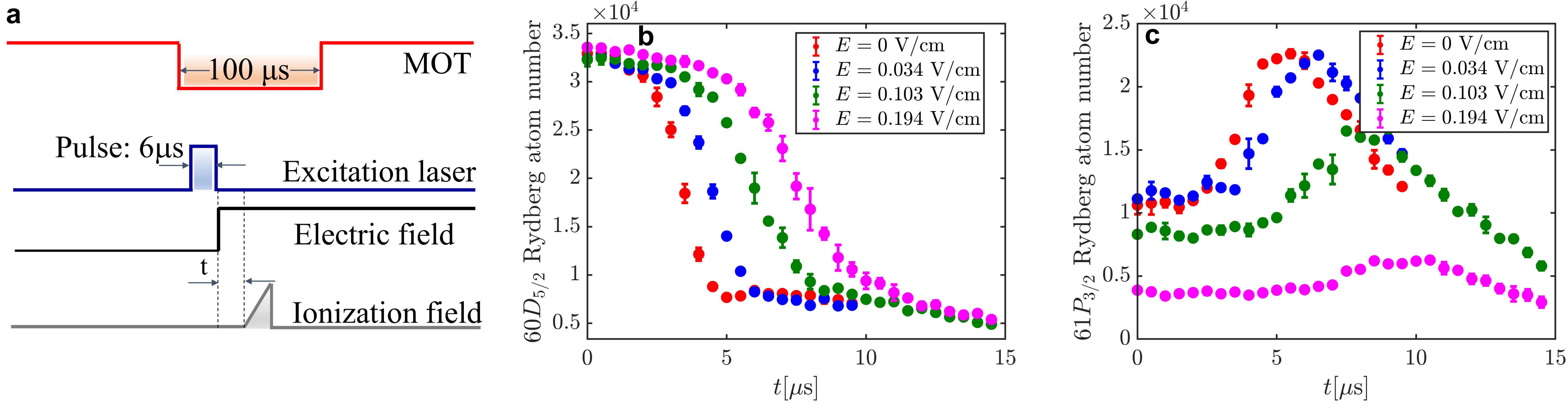}\\
	\caption{\footnotesize(Colour online) \textbf{Dipole-dipole interactions}. (a) A static electric field is applied immediately after preparing the initial states, which aligns the dipoles, leading to non-negligible dipole-dipole interactions. (b) Population in state $|60D_{5/2}\rangle$. Superradiant dynamics slows down when the dipole-dipole interaction is  present. The stronger the electric field, the slower the decay of the population. The number of Rydberg atoms is same for different $E$ field. (c) Population in state $|61P_{3/2}\rangle$. Due to the dipole-dipole interaction, the population in state $|61P_{3/2}\rangle$ grows slower when the electric field is stronger.
	}\label{Efield}
\end{figure*}

To further identify electric field effects on the Rydberg transitions, we applied a microwave electric (MW) field with frequency $3.21573$\,GHz resonantly interacting with the $|60D_{5/2}\rangle\to|61P_{3/2}\rangle$ transition, as shown in Fig.~\ref{MW}(a). In the presence of the MW field, the TOF signals are changed apparently, i.e. more Rydberg atoms are transferred to the $|61P\rangle$ state, see Fig.~\ref{MW}(b). It can be seen that the TOF signals in the presence of the MW field appear at the same position as the one without applying the MW field, indicating that the signal is indeed due to the $|60D_{5/2}\rangle\to|61P_{3/2}\rangle$ decay.   This change of populations is a result of the interplay between the MW field and superradiant decay.  To highlight the effect of the MW coupling, we show the population difference, i.e. subtracting the ion signal when the MW field  is off (bottom) from the one (top) with the MW field. As shown in Fig.~\ref{MW}(c), the population in state $|60D\rangle$ reduces (blue gate), and population in state $|61P\rangle$ states increases (red gate), due to the MW coupling. The profile demonstrates  that the superradiant decay takes place between these two states.  Another feature in the presence of the MW field is that the delay time of the TOF signal is slightly increased, possibly due to the interplay between the dipolar interactions and superradiance.

On the other hand, it is known that dipole-dipole  interactions will depend on angle $\theta_{jk}$ between the dipole and molecule axis that connects the $j$- and $k$-th Rydberg atoms, i.e.  $V_{jk}^e\propto C_3[1-3{{\rm cos}^2(\theta_{jk})}]/R_{jk}^3$. However without external fields, there are a large number of Rydberg atoms, such that the net interaction for any Rydberg atoms vanishes, i.e. $\sum_{k}V_{jk} \propto \int_0^{\pi} (1-3\cos^2\theta)\sin\theta d\theta=0$, where we have replaced the sum by a continuous integral in the estimation.

To check this, we have carried out additional experiments by applying a static electric field to the sample. See the timing and results in Fig.~\ref{Efield}. The electric field will align the dipoles along the direction of the field. In this case, a net dipole-dipole interaction will be induced. The presence of dipole-dipole interactions will slow down the superradiance due to many-body  dephasing~\cite{gross_superradiance:_1982}. In our experiment, we indeed find that speed of the superradiant decay is reduced when the electric field is strong, see Fig.~\ref{Efield}(b)-(c), where stronger electric fields give stronger dipole-dipole interactions, and hence cause slower superradiant decay.

\section{Conclusion and discussion}
We have observed the superradiant decay of the $|nD\rangle \to |(n+1)P\rangle$ transition in an ensemble of laser-cooled caesium Rydberg atoms in free space. The superradiance is found to be enhanced by finite temperature BBR due to high number densities of MW photons, confirmed by many-body simulations. The vdW interaction drastically modifies superradiance, leading to state dependent scaling.

Our system offers a controllable platform to investigate the interplay between strong collective dissipation and two-body Rydberg interactions. For example, superradiant dynamics will be qualitatively different at longer times when vdW interactions are strong. Dipole-dipole interactions, induced by external fields, will compete with the collective dissipation, too. Such competition can be explored experimentally in a controllable fashion. Beyond fundamental interests, our study might be useful in developing BBR thermometry in the MW domain whose sensitivities can be improved by collective light-atom interactions, with applications to improve accuracy of atomic clocks~\cite{Levi_Cryogenic_2010, Middelmann_Blackbody_2011, Ovsiannikov_Clocks_2011, Ushijima_clocks_2015}.

\begin{acknowledgements}
The work is supported by the National Key R\&D Program of China (Grant No.~2017YFA0304203), the National Natural Science Foundation of China (Grant Nos.~61775124, 11804202, and 61835007). Z. B. and G. H. acknowledge the support from the National Natural Science Foundation of China (Grant Nos.~11904104 and 11975098), the Shanghai Sailing Program (Grant No.~18YF1407100), and the International Postdoctoral Exchange Fellowship Program (Grant No.~20180040). W. L. acknowledges support from the EPSRC through Grant No. EP/R04340X/1 via the QuantERA project “ERyQSenS”, the UKIERI-UGC Thematic Partnership (IND/CONT/G/16-17/73), and the Royal Society through the International Exchanges Cost Share award No. IEC$\backslash$NSFC$\backslash$181078. We are grateful for access to the Augusta High Performance Computing Facility at the University of Nottingham.
\end{acknowledgements}

\appendix
\section{Preparation and detection of Rydberg atoms in the experiment}\label{exp}
\begin{figure}
	\centering
	\includegraphics[width=0.4\textwidth]{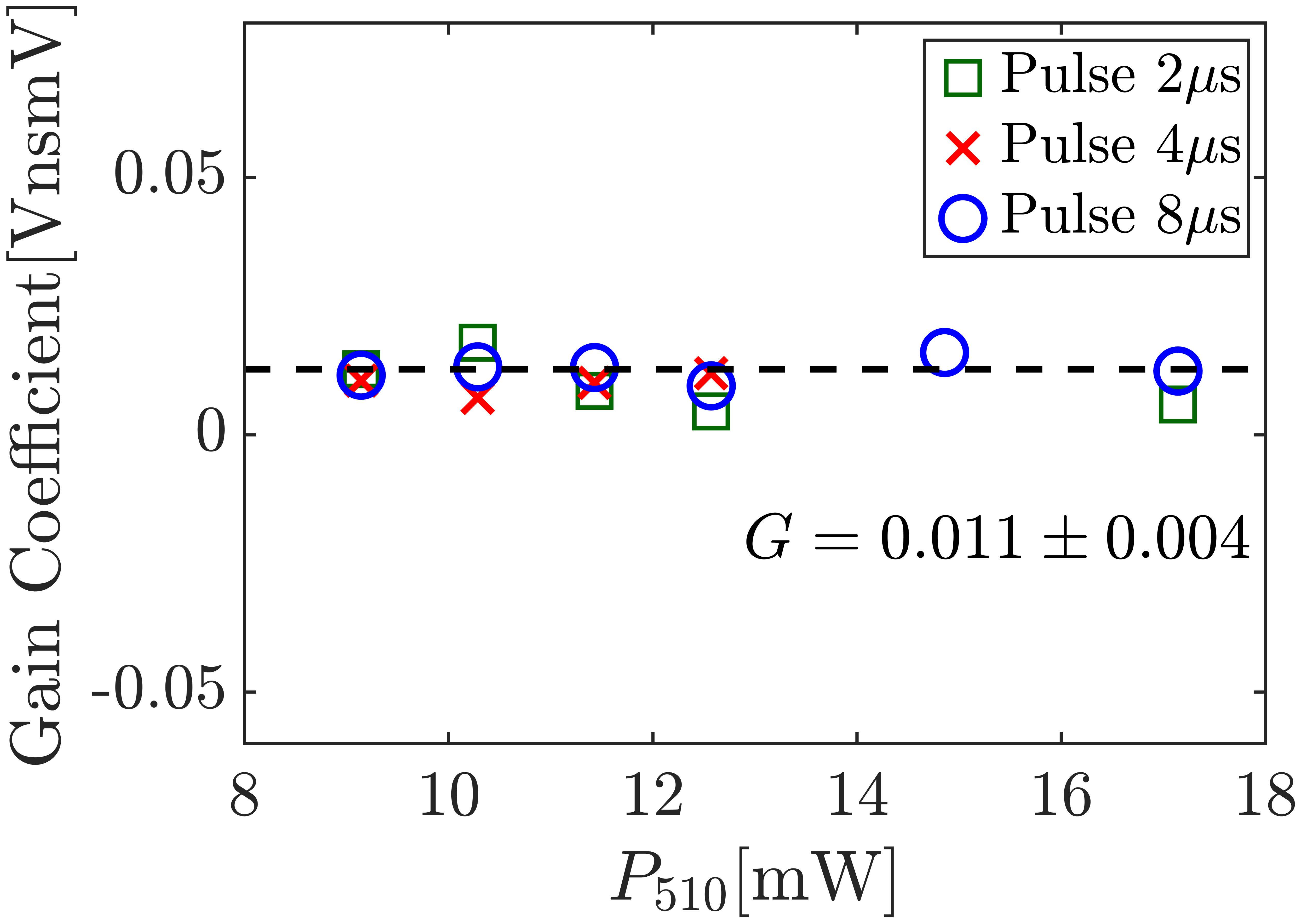}\\
	\caption{\footnotesize(Colour online) \textbf{Measured $G$ coefficient versus the power of coupling laser $P_{510}$}. Three fixed excitation pulses 2 $\mu$s (green square), 4 $\mu$s (red cross) and 8 $\mu$s (blue circle) are shown, respectively. The averaged value is shown as the dashed line.}\label{fig:gain}
\end{figure}
Our experiment is performed in a standard caesium magneto-optical trap (MOT). The atomic cloud has a diameter $\sim$550~$\mu$m and temperature $\sim$ 100~$\mu$K. After switching off the MOT beams, we apply a two-photon excitation lasers of 6~$\mu$s pulse to pump the groundstate atoms to Rydberg state $|nD_{5/2}\rangle$. The
state preparation process is described as follows: the probe laser (852~nm laser, Toptica DLpro) drives ground $|6S_{1/2}, F = 4\rangle$ $\to$ $|6P_{3/2},  F' = 5 \rangle$ transition and the coupling laser (510~nm laser, Toptica TA-SHG110) couples the $|6P_{3/2}, F'=5 \rangle$ $\to$ $|nD_{5/2}\rangle$ transition.
The laser frequencies are stabilized using a super stable optical cavity with 1.5-GHz FSR and 15000 fineness and the 852~nm laser is blue shifted 360~MHz from the intermediate level $|6P_{3/2}, F'=5\rangle$ using a double-pass acousto-optic modulator (AOM). The 852 and 510 nm beams, with respective waist of 80~$\mu$m and 40~$\mu$m, are overlapped at the MOT center in a counter propagating geometry yielding a cylindrical excitation region. Typical Rabi frequencies of the two lasers are $\Omega_p = 2\pi\times 132.05$\,MHz and $\Omega_c=2\pi\times 6.91$\,MHz. The excitation region is surrounded by three pairs of field-compensation electrodes, which allow us to reduce stray electric fields via Stark spectroscopy, corresponding stray field less than 30~mV/cm.

To experimentally measure Rydberg population, we use a ramping electric field to ionize the Rydberg atoms. In the experiments, the electric field is linearly increased to 256 V/cm in 3 $\mu$s, unless stated elsewhere explicitly. This field is much larger than the ionization threshold. Resultant ions are detected with a micro-channel plate (MCP) detector with a detection efficiency $\sim$ 10\%. The detected ion signals are amplified with an amplifier and analyzed with a boxcar integrator (SR250) and then recorded with a computer.

Before measuring Rydberg atoms, we first calibrate the MCP ions detection system with two shadow images taken before and after the laser excitation. From the difference of two shadow images, we obtain the number of Rydberg excitation, $N_e$, and therefore the gain factor, $G$, of the MCP ions detection system. The gain factor is defined as,
\begin{equation}\label{SM_G2}
G = \frac{V_{\rm signal}\cdot t_{\rm gate}\cdot S_{\rm sensitivity}}{N_e},
\end{equation}
where $V_{\rm signal}$ is the intensity of the measured ion signal, $t_{\rm gate}$ is the Boxcar gate width, and $S_{\rm sensitivity}$ is the Boxcar setting sensitivity, respectively. The shadow image is usually used to detect the number of MOT atoms. From the difference of two shadow images taken before and after the Rydberg excitation, we can extract the number of Rydberg excitation, which is compared to the ionization signal of Rydberg atoms, $V_{\rm signal}$. Using Eq.~(\ref{SM_G2}), we  obtain the gain factor, shown in Fig.~\ref{fig:gain}. For different Rydberg excitation power and pulse duration, the averaged gain factor is $G$ = 0.011$\pm$ 0.004. Considering the detection efficiency (10\%), the effective gain factor of the MCP detection system is $G_{\rm eff}$ = 0.11 $ \pm$ 0.04 in our experiment, which is used throughout the experiment to determine the number of Rydberg atoms.

\begin{figure}
	\centering
	\includegraphics[width=0.9\linewidth]{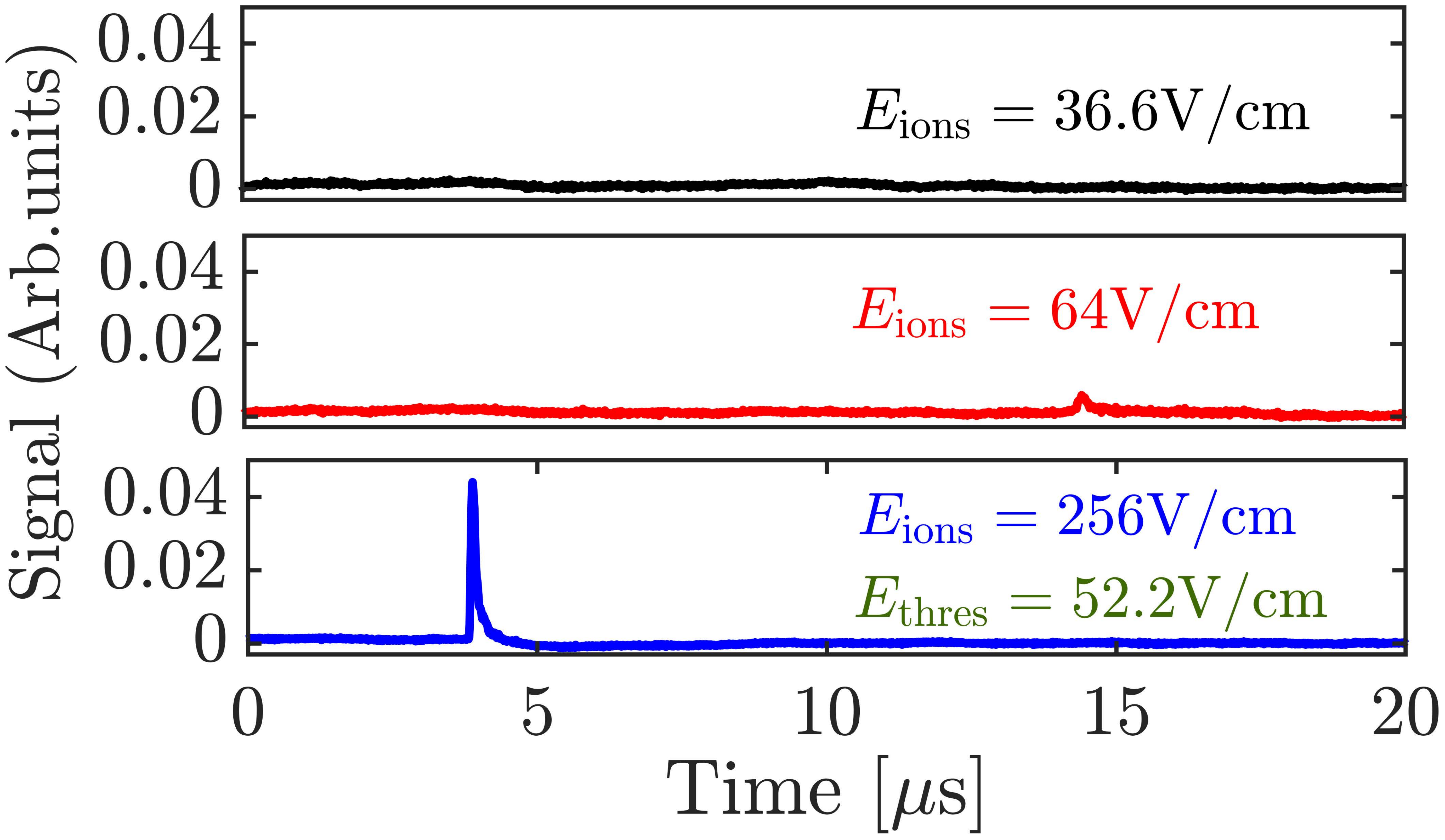}
	\caption{\footnotesize(Colour online) \textbf{Time of flight spectroscopy of field ionization of Rydberg atoms}. The black line represents for $E_{\rm ion} < E_{\rm thres}=52.2\,{\rm V/cm}$, and  the red line represents for $E_{\rm ion} > E_{\rm thres}$. When the ionization field is low, Rydberg ion signals are negligible, indicating that auto-ionization of Rydberg atoms is not present in the experiment. The Rydberg atoms are ionized only when $E_{\rm ion} > E_{\rm thres}$. Specially, when the ionization field $E_{\rm ion}=256$\,V/cm (blue line), Rydberg ion signal becomes strong.  Here we consider the Rydberg state $|\uparrow\rangle=|60D_{5/2}\rangle$.}
	\label{fig:tof}
\end{figure}

To verify our experimental signals is the field ionization of Rydberg atoms instead of the free ion signal, we have made experimental tests, with the result illustrated in Fig.~\ref{fig:tof}.
It is seen that the Rydberg population can be detected when the ionization field is larger than the ionization threshold of respective Rydberg states. When the electric field is lower than the threshold field, ion signals vanish (except fluctuations coming from background noise). This makes sure that the signal we measured is due to decay of Rydberg states, i.e. excluding auto-ionization of Rydberg atoms~\cite{Tanner_Ionization_2008,Viteau_Melting_2008}.

\section{Many-body dynamics based on the four-level model}\label{4LS}
The two-photon excitation is used to pump the groundstate atoms to the initial state $|nD_{5/2}\rangle$ in the experiment. In the experiment, four states (two lower and two Rydberg states) are involved. Here, we denote $|1\rangle=|6S\rangle$, $|2\rangle=|6P\rangle$,  $|3\rangle=|nD\rangle$, and $|4\rangle=|(n+1)P\rangle$. The system can be described by the Hamiltonian $\hat{H}=\hat{H}_a + \hat{H}_{int}$ in the interaction picture and rotating-wave approximation $(\hbar=1)$,
\begin{subequations}\label{Hamiltonian1}
\begin{eqnarray}
& & \hat{H}_a=
\sum_{j=1}^{N}\left[-\Delta_2 \hat{\sigma}_{22}^j+\Omega_p\hat{\sigma}_{21}^j(\mathbf{r},t)
+\Omega_c\hat{\sigma}_{32}^j(\mathbf{r},t)+{\rm H.c.}\right],\nonumber\\
& &\\
& & \hat{H}_{int}=
\sum_{j=1}^N\sum_{k\neq j}^N\left[\frac{1}{2}V_{jk}^{33}\hat{\sigma}_{33}^j\hat{\sigma}_{33}^k+\frac{1}{2}V_{jk}^{44}\hat{\sigma}_{44}^j\hat{\sigma}_{44}^k\right],
\end{eqnarray}
\end{subequations}
where $\hat{\sigma}_{\alpha\beta}^j=|\alpha\rangle\langle\beta|_j$ ($\alpha,\beta=1, 2, 3, 4$) is the transition operator of the $j$th atom. The dissipation effect is described by the Lindblad operator $D_1(\rho)$ and $D_2(\rho)$,
\begin{subequations}\label{Lindblad2}
\begin{eqnarray}
D_1(\rho)=&&\sum_{j,k}\Gamma_{jk}\left(\hat{\sigma}_{43}^j\rho\hat{\sigma}_{34}^k-\frac{1}{2}\{\hat{\sigma}_{34}^k\hat{\sigma}_{43}^j, \rho\}\right),\\
D_2(\rho)=&&\sum_{j=1}^N\Gamma_{12}\left(\hat{\sigma}_{12}^j\rho\hat{\sigma}_{21}^j-\frac{1}{2}\{\hat{\sigma}_{22}^j, \rho\}\right)\nonumber\\
&&+\Gamma_{23}\left(\hat{\sigma}_{23}^j\rho\hat{\sigma}_{32}^j-\frac{1}{2}\{\hat{\sigma}_{33}^j, \rho\}\right),
\end{eqnarray}
\end{subequations}
where $D_1$ denotes the collective radiation between the Rydberg states and $D_2$ describes the single-body decay between state $|2\rangle(|3\rangle)$ to state $|1\rangle(|4\rangle)$ with rate $\Gamma_{12}$ ($\Gamma_{34}$).

For a few particles, the master equation can be solved numerically. To capture the build up of superradiant emission in a large ensemble, we employ a generalized discrete truncated Wigner approximation (GDTWA) based on a
Monte Carlo sampling in phase space, where GDTWA method can effectively capture complex quantum dynamics in high spin systems~\cite{lepoutre_out--equilibrium_2019}. The generic density matrix for a discrete system with $D$ states takes the form $\hat{\rho}_i=\sum_{\alpha=1,\beta=1}^D c_{\alpha\beta}|\alpha\rangle\langle\beta|$. For $D=4$ (equivalent to a spin-3/2 atom), the states $|\alpha\rangle$ with $\alpha=1, 2, 3, 4$ associates to the spin states $m_s=-3/2, -1/2, 1/2, 3/2$ of the spin-3/2 atom. Since $(\hat{\rho}_i)^\dagger=\hat{\rho}_i$ and Tr$(\hat{\rho}_i)=1$ and  total $(D^2-1)$ real numbers are needed to describe an arbitrary state, which can be expressed as average values of $(D^2-1)$ orthogonal observable:
\begin{widetext}
\begin{eqnarray}
&\hat{\Lambda}_{\alpha, \beta<\alpha}^{[i],R}=(|\beta\rangle\langle\alpha|+|\alpha\rangle\langle\beta|),~~~~~ 1\leq\alpha\leq D, 1\leq\beta\leq D-1\nonumber\\
&\hat{\Lambda}_{\alpha, \beta<\alpha}^{[i],I}=-i(|\beta\rangle\langle\alpha|-|\alpha\rangle\langle\beta|),~~~~~ 1\leq\alpha\leq D, 1\leq\beta\leq D-1\nonumber\\
&\hat{\Lambda}_{\alpha}^{[i],D}=\sqrt{\frac{2}{\alpha(\alpha+1)}}\left(\sum_{\beta=1}^\alpha|\beta\rangle\langle\beta|-\alpha|\alpha+1\rangle\langle\alpha+1|\right),~~~~~ 1\leq\alpha\leq D-1,
\end{eqnarray}
\end{widetext}
where $\hat{\Lambda}_{\alpha, \beta<\alpha}^{[i],R/I}$ correspond to the real (``R'') and imaginary (``I'') parts of the off-diagonal parts of $c_{\alpha\beta}$ and $\hat{\Lambda}_{\alpha}^{[i],D}$ to linear combinations of the real diagonal elements $c_{\alpha\alpha}$. Note that for $D=2$, the matrices are standard Pauli matrices for spin $1/2$ system (see the DTWA method in main text). For $D>2$, the matrices reduce to a generalized Gell-Mann matrices (GGMs) and corresponds to SU($D$) group for spin-$(D-1)/2$ system.

In the GDTWA method, we describe the initial state by a probability ``Wigner'' distribution, $p_{\mu,a_\mu}^{[i]}$ with $a_\mu$ denoting the index of each trajectory~\cite{lepoutre_out--equilibrium_2019}. The discrete set of initial configurations, $\{ \lambda_{\mu}^{[i]}\}$, can be interpreted by a ``projective measurement of the GGM'': for each $\lambda_{\mu}^{[i]}$, we
choose a set of initial configurations given by the eigenvalues of each GGM. Consider the eigen-expansion of the GGMs,
\begin{eqnarray}
\hat{\Lambda}_{\mu}^{[i]}=\sum_{a_\mu}\eta_{\mu,a_\mu}^{[i]}|\eta_{\mu,a_\mu}^{[i]}\rangle\langle\eta_{\mu,a_\mu}^{[i]}|,
\end{eqnarray}
where $\eta_{\mu,a_\mu}^{[i]}$ and $|\eta_{\mu,a_\mu}^{[i]}\rangle$ denote the eigenvalues and eigenvectors, respectively. Then, we select the initial condition $\lambda_{\mu}^{[i]}(t=0)=\eta_{\mu,a_\mu}^{[i]}/2$, with probability $p_{\mu,a_\mu}^{[i]}={\rm tr}[\hat{\rho}_0^{[i]}|\eta_{\mu,a_\mu}^{[i]}\rangle\langle\eta_{\mu,a_\mu}^{[i]}|]$. Specifically, for the initial state
\begin{eqnarray}
&\hat{\rho}_0^{[i]}=|1\rangle\langle1|=\left(
                                               \begin{array}{cccc}
                                                 1 & 0 & 0 & 0 \\
                                                 0 & 0 & 0 & 0 \\
                                                 0 & 0 & 0 & 0 \\
                                                 0 & 0 & 0 & 0 \\
                                               \end{array}
                                             \right),\nonumber
\end{eqnarray}
which leads to fixed diagonal Bloch vector element $\lambda_{1}^{[i],D}=1/2$, $\lambda_{2}^{[i],D}=1/\sqrt{12}$, $\lambda_{2}^{[i],D}=1/\sqrt{24}$, fixed off-diagonal elements $\lambda_{\alpha>\beta, \beta>1}^{[i],D}=0$ and fluctuating off-diagonal elements $\lambda_{\alpha=1,2,3,4, \beta=1}^{[i],D}\in\{-1/2, 1/2\}$, each with $50\%$ probability.
\begin{figure}
\centering
\includegraphics[width=1\linewidth]{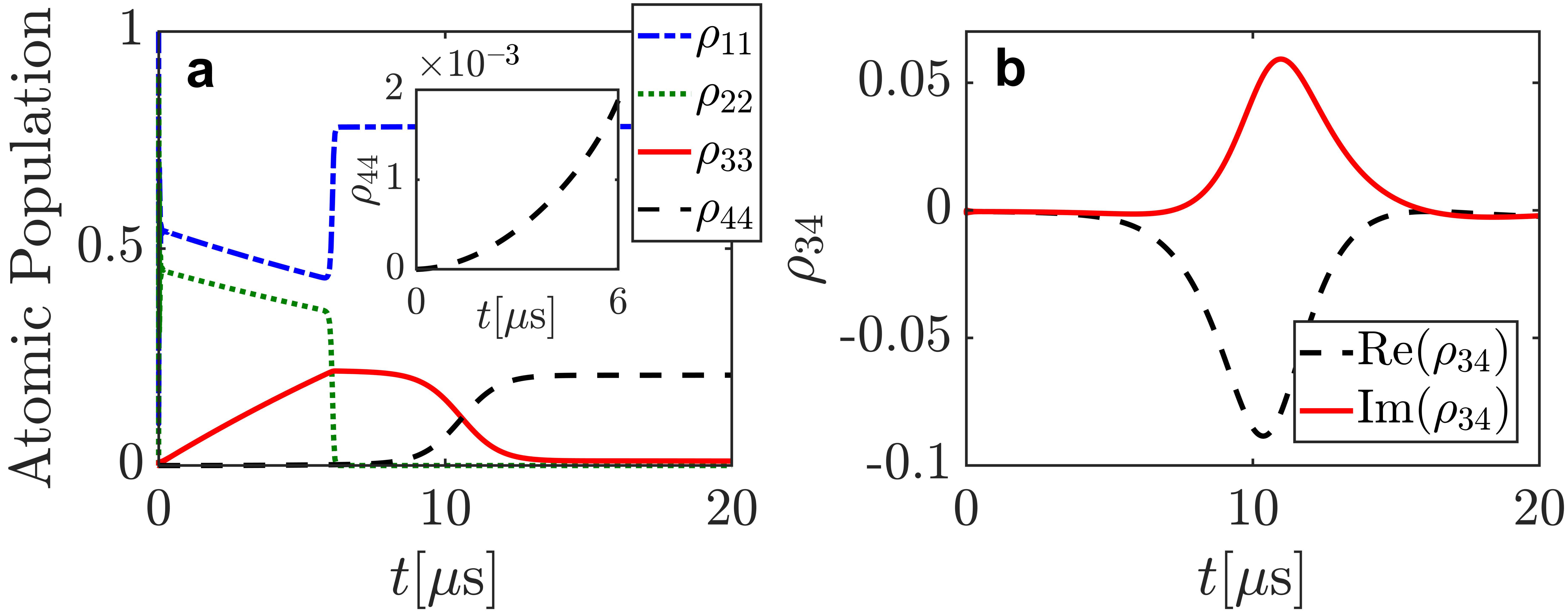}\\
\caption{\footnotesize(Colour online) \textbf{GDTWA simulation of the 4-level model}. (a) Time evolution of atomic populations of levels $|1\rangle$, $|2\rangle$, $|3\rangle$ and $|4\rangle$. All atoms initially populate at the groundstate $\Pi_{j=1}^N|1_j\rangle$, and the probe and control fields are switched off at $t=6~{\rm \mu s}$. Before the excitation laser is turned off, the population in state $|4\rangle$ is very small (inset). After the laser is turned off, a fast population transfer from state $|3\rangle$ to state $|4\rangle$ is found.  (b) The time evolution of the real and imaginary part of coherence $\rho_{34}$ between Rydberg states for one trajectory. Im$(\rho_{34})$ has a  {\it sech} form due to superradiance.}\label{population2}
\end{figure}


To show dynamics starting from the laser excitation, we have made simulations with the following parameters: $\Gamma=389.9$\,Hz ($n=60$),  the number for groundstate atom $N=6000$, $\Gamma_{12}=2\pi\times5.2\,{\rm MHz}$, $\Omega_p=2\pi\times132.05\,{\rm MHz}$, $\Omega_c=2\pi\times6.91\,{\rm MHz}$, and $\Delta=360\,{\rm MHz}$. Here $N$ is the number of groundstate atoms (not the number of Rydberg state atoms). Here superradiance takes place on a much longer time scale, as the number of atoms can be excited to Rydberg states is small. To mimic the experiment, we have increased the single-body decay rate by a factor of 3, in order to illustrate the $|3\rangle \to |4\rangle$ decay. As shown in Fig.~\ref{population2}(a), about $22\%$ atoms are excited to state $|3\rangle$ during the laser excitation. In the mean time state $|4\rangle$ is populated weakly, which is seen in the experiment. Once the laser is turned off, superradiance is triggered. We see rapid population transfer $|3\rangle\to |4\rangle$ when $t>6\,\mu$s. When looking at the coherence $\rho_{34}$, we find that its profile shows a hyperbolic function form, due to the emergence of superradiance.

It is not possible for us to simulate system sizes close to the experiment with the 4-level model,  even using the GDTWA. In typical experiments hundreds of thousand atoms interact with laser fields. Among them, tens of thousand atoms are excited to Rydberg states. It is numerically challenging to simulate such large systems. As we focus on dynamics after the laser is switched off, this allows us to apply the two-level approximation. In this way, we can efficiently simulate dynamics of large system sizes by excluding groundstate atoms from the model.

%
%

\end{document}